\documentclass[12pt]{article}

\usepackage{amssymb,amsbsy,times,fancyhdr,color}

\usepackage{amsmath,latexsym}

\usepackage[dvips]{epsfig}

\usepackage{float,blkarray,accents}

\usepackage{multirow,subcaption}

\usepackage{tikz,booktabs,setspace}

\usepackage[round]{natbib}
\newcommand{\bA}{\mathbf{A} }

\newcommand{\bB}{\mathbf{B} }

\newcommand{\bD}{\mathbf{D} }

\newcommand{\bH}{\mathbf{H} } 

\newcommand{\bI}{\mathbf{I} }

\newcommand{\bR}{\mathbf{R} }

\newcommand{\bX}{\mathbf{X} }

\newcommand{\half}{\frac{1}{2}}

\newcommand{\vect}[1]{\underaccent{\bar}{#1}}

\newcommand{\nn}{\nonumber}

\newcommand{\beq}{\begin{equation}}
\newcommand{\eeq}{\end{equation}}

\begin{document}
\title{\sc {\bf The Classical Multidimensional Scaling Revisited}}
\author{Kanti V. Mardia, \\
 University of Leeds and University of Oxford,\\ and \\
Anthony D. Riley, \\ 
University of Leeds.}
 
\date{ }
\maketitle
\section*{Abstract}
We reexamine the the classical multidimensional scaling (MDS). We study  some special cases, in particular,  the exact solution for  the sub-space formed by the  3 dimensional principal coordinates is derived. Also we give the extreme case when  the points are collinear.  Some insight into the effect on the  MDS solution   of  the excluded  eigenvalues (could be both positive as well as  negative) of the doubly centered matrix is provided. As an illustration, we work through an example  to understand the  distortion in the MDS construction with positive and negative eigenvalues.

\section{Basics of the classical MDS} 
We  recall  in this section  some basics  of the classical MDS from  Mardia et al (1979, Section 14.2).
Let us denote the $n\times n$ distance matrix as $\bD =(d_{ij})$ and form  the matrix $\bA$
\begin{align}
\bA = \{a_{ij}\}, a_{ij} &= -\frac{1}{2}d_{ij}^2,
\end{align}
and define the corresponding  doubly centered matrix $\bB$,
\beq
\bB = \bH\bA\bH, 
\eeq
where $\bH = \mathbf{I_n}-\frac{\vect{1}_n \vect{1}_n^T}{n}$ is the centering matrix in the standard notation. We can rewrite $\bB$ as 
\beq
\bB = \bX\bX^T,
\eeq
and  the  $n \times n$ matrix $\bX$ of  the principal coordinates in the Euclidean space  
 (assuming that  $\bB$ is semi-positive) is given by 
\beq \label{X}
\bX =\mathbf{\Gamma}\mathbf{\Lambda}^{\half}=({\lambda_1}^{\frac{1}{2}}{\gamma_1},{\lambda_2}^{\frac{1}{2}}{\gamma_2}\ldots, {\lambda_n}^{\frac{1}{2}}{\gamma_n})= (\vect{x}_{(1)},\vect{x}_{(2)}, \ldots, \vect{x}_{(n)}),
\eeq
where  
$$\vect{x}_{(i)}={\lambda_i}^{\frac{1}{2}}{\vect{\gamma}_i}, i=1,\ldots,n ,$$
with   the spectral decomposition of $\bB$ 
\beq \label{Bspectral}
\bB = \mathbf{\Gamma} \mathbf{\Lambda} \mathbf{\Gamma}^T,
\eeq
and $$\mathbf{\Gamma}= \{\gamma_{ij}\}=(\vect{\gamma}_1,\ldots, \vect{\gamma}_n)$$ 
is the orthogonal matrix of eigenvectors and $\mathbf{\Lambda}$ is the diagonal matrix of eigenvalues,
$$\mathbf{\Lambda} = \text{Diag}\{\lambda_{1},\ldots,\lambda_{n} \}.$$
Indeed, the $n$ principal coordinates $\vect{x}_i, i=1,2, \ldots,n $ are the rows of $\bX$, namely,
\beq \label{Xcoord}
\bX=
\begin{pmatrix}
   \vect{x}^T_1  \\
   \vect{x}^T_2 \\
    \vdots  \\
    \vect{x}^T_n
\end{pmatrix}
, \vect{x}_i^T=({\lambda_1}^{\frac{1}{2}}{\gamma}_{i1},\ldots,{\lambda_n}^{\frac{1}{2}}{\gamma}_{in}), i=1,\ldots, n. 
\eeq
We can  use any ``subpart" of $\bX$ to define the principal coordinates of a low dimensional space as our MDS solution.  Note that the last eigenvalue $\lambda_{n}$ is zero so at least $\vect{x}_{(n)}=0$ so we can work on the remaining $n-1$ dimensional  coordinates. 

Note that, for simplicity, we have taken $\bX$ as the $n \times n$ matrix rather than $n \times p$ 
matrix.  We now show that, for any dissimilarity  matrix $\bD$ with real entries but not necessarily semi-positive definite  $\bB$ as in above, $\lambda_1$ will be  always positive. We have  
$$ tr(\bB) = tr (\bH^2\bA)=tr (\bH\bA)= tr (\bA)-tr(\vect{1}^T \bA \vect{1})/n$$
so that 
\beq \label{tr}
\sum_{i=1}^{n} \lambda_i = \sum_{i<j}d_{i,j}^2/n >0.
\eeq
Hence $\lambda_1>0$. Thus implying that  we can always "fit" one-dimensional configuration for any distance/ dissimilarity matrix.  

Let us now consider the case when the n points lie on a line then we will have only one non-zero eigenvalue $\lambda$ of $\bB$ so from \eqref{tr}, it is given by 
$$\lambda = \sum_{i<j}d_{i,j}^2/n. $$
\begin{itemize}
\item For n=3 with points on the line with the inter-point distances as $a,b,c$, we have 
$\lambda =(a^2+b^2+c^2)/3,$ where if $AB=a, BC=b,AC=c$ with the points $A,B,C$ in that order then $a+b=c.$
\item If the points are  $1,2,\ldots,n$  then we find that
 $\lambda = n(n^2-1)/12$
so with the distances scaled to (0,1), we have $\lambda$ = $n(n^2-1)/12(n-1)^2$ and $\lambda$=O(n). 
\end{itemize}
\section{The MDS solution for $2 \times 2$ distance matrix}
Let $X=(\vect{x}_{(1)},\vect{x}_{(2)})$ where
 $\vect{x}_k,k=1,2$ are the  coordinates in 2 dimensions. Suppose    the two points are separated by a distance $d$.
 We have
\begin{align}
D = 
\begin{pmatrix}
 0 & d \\
 d & 0
\end{pmatrix} , \hspace{3pt}
B =
\begin{pmatrix}
  \frac{d^{2}}{4} & \frac{-d^{2}}{4} \\
  \frac{-d^{2}}{4} & \frac{d^{2}}{4}
\end{pmatrix}. \nn
\intertext{The eigenvalues of $\bB$ are given by}
|\bB-\lambda\bI_{2}| &= \lambda^{2}-2 \lambda \frac{d^{2}}{4}. \label{DecompersionTwo}
\intertext{Solving \eqref{DecompersionTwo}  gives the  eigenvalues}
\lambda_{1} = \frac{d^{2}}{2} \ &\text{and} \ \lambda_{2} = 0, \nn
& \nn
\intertext{with the corresponding eigenvectors  }
\vect{\gamma}_{1} = \left(\frac{1}{\sqrt{2}}, \frac{-1}{\sqrt{2}}\right)^{T} 
\ &\text{and} \ 
\vect{\gamma}_{2}  = \left(\frac{1}{\sqrt{2}}, \frac{1}{\sqrt{2}}\right)^{T}. \nn
\intertext{Now using \hspace{3pt}$\vect{x}_{(k)}=\lambda_{k}^{\frac{1}{2}}\vect{\gamma}_{k}$ from   \eqref{X}, and we get for $\bX $}
\vect{x}_{(1)}  = \left(\frac{d}{2}, \frac{-d}{2}\right)^{T} \hspace{3pt} &\text{and} \hspace{3pt}
\vect{x}_{(2)}  = \left(0,0 \right)^{T} \label{TwoSol}.
\end{align}
So the principal coordinates from \eqref{Xcoord} are 
\beq \label{coord2D}
\vect{x}_1  = \left(\frac{d}{2}, 0\right)^{T} \hspace{3pt} \text{and} \hspace{3pt}
\vect{x}_2  = \left(-\frac{d}{2},0 \right)^{T}.
\eeq
  To get   a lower dimensional coordinate space (one dimensional), we can simply use, along the line ( $x$- axis),  the following  two points:
$$x_1=\frac{d}{2},\quad x_2=-\frac{d}{2}.$$
We can now shift  $x $ conveniently  by using $ x^*$ = $ x +\frac{d}{2} $
 so we have the new coordinates  $$x^*_1=d,\quad x^*_2=0 $$ along the $x^*$- axis in one dimension with the origin at $x^*_2$.
 
The solution \eqref{coord2D} is trivial as the points only require placing a distance $d$ apart to be recovered, although it does serve as a pointer for the $3 \times 3$ distance matrix in the next section. 
\section{The MDS solution for $3 \times 3$ distance matrix}
We now extend the last section of the $2 \times 2$ distance matrix to the $3 \times 3$ distance matrix where in principle, we need to follow the same steps. 
 Let now $\bX=(\vect{x}_{(1)},\vect{x}_{(2)},\vect{x}_{(3)})$, where $\vect{x}_k, k=1,2,3$ give the coordinates of points in three dimensions . Let
\begin{align}
\bD &=
\begin{pmatrix}
0 & a & b \\
a & 0 & c \\
b & c & 0
\end{pmatrix}.\\  \text{Then it can be seen that} \nn \\
\bB & = \frac{1}{18}
\begin{pmatrix}
4a^2+4b^2-2c^2 & -5a^2+b^2+c^2 & a^2-5b^2+c^2 \\
-5a^2+b^2+c^2 & 4a^2-2b^2+4c^2 & a^2+b^2-5c^2 \\
a^2-5b^2+c^2 & a^2+b^2-5c^2 & -2a^2+4b^2+4c^2 \end{pmatrix}. \nn
\intertext{and}
|\bB-\lambda\bI_{3}| &= \left(\frac{-1}{6}(a^{2}b^{2}+a^{2}c^{2}+b^{2}c^{2}) +\frac{1}{12}(a^{4}+b^{4}+c^{4})\right)\lambda \nn \\
                     & \quad \quad  +\frac{1}{3}(a^{2}+b^{2}+c^{2})\lambda^{2}-\lambda^{3}. \label{DecompersionThree}\\
\intertext{Let}                     
\Delta&=\sqrt{a^{4}+b^{4}+c^{4}-a^{2}b^{2}-a^{2}c^{2}-b^{2}c^{2}} \label{Delta}.\\     
 \intertext{Solving \eqref{DecompersionThree} we find that   the eigenvalues are }                   
\lambda_1 &=\frac{1}{6}(a^{2}+b^{2}+c^{2} +2\Delta ), \nn \\
\lambda_2 &=\frac{1}{6}(a^{2}+b^{2}+c^{2} -2\Delta ), \nn \\
\text{and} \ \lambda_3 &=0. \label{EigenvaluesThree}
\intertext{We show below in the proof of Theorem 1 that $\Delta$ is always non-negative. 
To find the corresponding eigenvectors, $\bB$ is rotated using a Helmert rotation matrix $\bR$}
\bR &= \begin{pmatrix}
\frac{1}{\sqrt{3}} & \frac{1}{\sqrt{3}} & \frac{1}{\sqrt{3}} \\
\frac{-1}{\sqrt{2}} & 0 & \frac{1}{\sqrt{2}} \\
\frac{1}{\sqrt{6}} & \frac{-2}{\sqrt{6}} & \frac{1}{\sqrt{6}}
\end{pmatrix}. \nn \\
\intertext{That is}
\bR\bB\bR^{T} &=
\begin{pmatrix}
0 & 0 & 0 \\
0 & \frac{b^2}{2} & \frac{(-a^2+c^2)}{2\sqrt{3}} \\
0 & \frac{(-a^2+c^2)}{2\sqrt{3}} & \frac{(2a^2 - b^2 +2c^2)}{6}
\end{pmatrix}
=\begin{pmatrix}
0 & 0 & 0 \\
0 & \sigma^{2}_{1,1} & \sigma^{2}_{1,2} \\
0 & \sigma^{2}_{1,2} & \sigma^{2}_{2,2}
\end{pmatrix} \label{NullMat1}
\end{align}
which has a $2 \times 2$ symmetric matrix nested within a $3 \times 3$ null matrix. We first  give the 
eigenvectors of $\bR^{T}\bB\bR$  by using a result of  Mardia et al (1979, page 246, Exercise 8.1.1) ,
\begin{align}
\vect{\phi}_1 &=
\begin{pmatrix}
0 \\
\sigma^{2}_{2,2}-\sigma^{2}_{1,1}+\Theta \\
-2\sigma^{2}_{1,2} 
\end{pmatrix} 
\hspace{3pt} \text{and} \hspace{3pt}
\vect{\phi}_2 =
\begin{pmatrix}
0 \\
2\sigma^{2}_{1,2} \\
\sigma^{2}_{2,2}-\sigma^{2}_{1,1}+\Theta
\end{pmatrix} \label{NullMatVect}
\intertext{where $\Theta = \sqrt{(\sigma^{2}_{1,1}-\sigma^{2}_{2,2})^{2}+4\sigma^4_{1,2}}$. Next, the rotation is reversed by pre-multiplying the  {\bf unnormalized} eigenvectors \eqref{NullMatVect} by $\bR$ to deduce the eigenvectors of $\bB$}
\vect{\gamma}_{1} &= 
\begin{pmatrix}
b^{2}-c^{2}+\Delta \\
-a^{2}+c^{2} \\
a^{2}-b^{2}-\Delta
\end{pmatrix},\ 
\vect{\gamma}_{2} =
\begin{pmatrix}
2a^{2}-b^{2}-c^{2}-\Delta \\
-a^{2}+2b^{2}-c^{2}+2\Delta \\
-a^{2}-b^{2}+2c^{2}-\Delta
\end{pmatrix}
, \ 
\vect{\gamma}_{3} =
\begin{pmatrix}
1 \\
1 \\
1 
\end{pmatrix} \label{EigenvectorsThree}
\intertext{ where $\Delta$ is given by  \eqref{Delta}. Now using \hspace{3pt}$\vect{x}_{(k)}=\lambda_{k}^{\frac{1}{2}}\vect{\gamma}_{k}$ from   \eqref{X}, and we get for $\bX $}
\vect{x}_{(1)} &= w_{1}  
\begin{pmatrix}
 b^{2}-c^{2}+\Delta \\
-a^{2}+c^{2} \\
a^{2}-b^{2} -\Delta
\end{pmatrix}  \label{x1}
\text{where} \nn \\
w_{1} &= \sqrt{(a^{2}+b^{2}+c^{2}+2 \Delta )/(12\delta )},\\
\delta &=\Delta (2\Delta-a^{2}+2 b^{2}-c^{2})\nn 
  \\
\intertext{and}
\vect{x}_{(2)} &= w_{2} 
\begin{pmatrix}
2a^{2}-b^{2}-c^{2}-\Delta \\
-a^{2}+2b^{2}-c^{2}+2\Delta \\
-a^{2}-b^{2}+2c^{2}-\Delta
\end{pmatrix} \label{x2}
\text{where}\nn \\
w_{2} &= 
\sqrt{(a^{2}+b^{2}+c^{2}-2 \Delta )/(12\delta )},  \\
\delta &=\Delta (2\Delta-a^{2}+2 b^{2}-c^{2})\nn \\
\intertext{and}
 \vect{x}_{(3)} &=(0,0,0)^{T} \label{x3}\\
 \text{since} \hspace{3pt} \lambda_{3}=0. \nn
\end{align}

The constants $w_{1}$ and $w_{2}$ are a product of $\lambda_{k}^{\frac{1}{2}}$ and the eigenvector normalization constant. 
 Let A, B and C be the vertices of the triangle then say  $AB=a, BC=b, AC=c $ .
Further, let $x,y,z $ be the three axes then  with $n=3$   the principal coordinates from \eqref{Xcoord} can be written down using $\vect{x}_{(1)},\vect{x}_{(2)},\vect{x}_{(3)}$ from \eqref{x1}, \eqref{x2}, and \eqref{x3} respectively.  As the triangle lies in the $x-y$ plane , we have the following theorem with $(x_i,y_i), i=1,2,3$ of A, B, C respectively by ignoring the $z-$ coordinates.\\
 {\bf Theorem 1.} 
Let $a^{2}+b^{2}+c^{2} \geq 2 \Delta $  where $\Delta$ is given by  \eqref{Delta} , we have 
   
 \beq \label {A}
 x_1=  w_{1} ( b^{2}-c^{2}+\Delta ),\quad y_1=  w_{2}(2a^{2}-b^{2}-c^{2}-\Delta ),
\eeq

\beq \label {B}
 x_2= w_{1} ( -a^{2}+c^{2} ),\quad y_2=  w_{2}(-a^{2}+2b^{2}-c^{2}+2\Delta  ),
\eeq

\beq \label {C}
 x_3= w_{1} (a^{2}-b^{2} -\Delta ),\quad  y_3=  w_{2}(-a^{2}-b^{2}+2c^{2}-\Delta  ),
\eeq
where  
$$w_{1} = \sqrt{(a^{2}+b^{2}+c^{2}+2 \Delta )/(12\delta )},\quad w_{2}=  
\sqrt{(a^{2}+b^{2}+c^{2}-2 \Delta )/(12\delta )},$$
with $\delta =\Delta (2\Delta-a^{2}+2 b^{2}-c^{2}).$  

Further, the center of gravity  of the triangle is at  $(0,0)$. 
\\
{\bf Proof.} Most of the results are already proved above. Note that $\Delta$ and $\delta$ are non-negative using the following inequality of  the Geometric Mean and Arithmetic Mean given below successively.
 $$ x^2y^2 \leq (x^4+y^4)/2.$$
 Alternatively, we see  easily   $ \Delta >0$ on noting that  
$$ 2\Delta ^2 = (a^2 -b^2)^2 + (a^2 -c^2)^2  + (b^2 -c^2)^2. $$
 
{\bf Corollary 1.} Let $a=c, a < b < 2a $ then for this isosceles case,  we have 
\beq \label {Iso} 
x_1= b/2, y_1= - (4a^{2}-b^{2})/6;\quad 
x_2= 0,y_2= -2 y_1 ;\quad 
x_3= -x_1,  y_3= y_1. 
\eeq
{\bf Proof.}  From the equations \eqref{EigenvaluesThree},  \eqref{x1}  and \eqref{x2}, we find    that   for  $a<b<2a$  we have   
$$\lambda_1 = b^2/2, \lambda_2 =(4 a^2 - b^2)/6, \quad  w_{1} = 1/(4\Delta ),\quad w_{2}=  
\sqrt{(4 a^{2}-b^{2})}/(12\Delta ),$$
where $\Delta =b^{2}-a^{2}.$ Using these results in  Theorem 1, our proof follows.  
\\ 
We now consider a wide range of particular isosceles triangles
\begin{itemize}
\item If $b=a$ , we have an equilateral triangle. 
\item If $b=2a$ , we have  a flat triangle as  $\lambda_2 = \lambda_3 =0.$ 
\item If $b>2a$ then $\lambda_1>0,  \lambda_3 =0$ but  $\lambda_2$ is imaginary so we can have a real solution only in one dimension.
\item If $a$ is very large and $b$ is fixed then we have a peaked isosceles triangle.
\end{itemize}

Note that for the isosceles triangle, without any loss of generalities by rescaling,  we can write the  coordinates of $A,B,C$ as 
$$ A=(1,-e),B=(0,2e),C=(-1,-e)$$ where $ e= \sqrt{(4 a^{2}-b^{2})}/(3 b).$
It allows the equilateral case with  $a=b$ ( as a limit) leading to  the coordinates 
$$ A(1,-1/\sqrt{3}),B(0,2/\sqrt{3}),C(-1,-1/\sqrt{3}).$$
 
{\bf Remark 1.} Equation \eqref{EigenvaluesThree}, which gives the eigenvalues of $\bB$, can be used to determine if the desired Euclidean properties of $\bD$ are violated. Rearranging the equation for the second eigenvalues \eqref{EigenvaluesThree}  or $w_2 \geq 0$ gives the condition ( for $\bB$ to be semi- positive definite)
\begin{align}
a^{2}+b^{2}+c^{2} &\geq 2 \Delta. 
\end{align}
Hence,  if this inequality holds then $\bD$ is Euclidean.

{\bf Remark 2.} For visualization, we can shift the origin (and rotate if so desired)  for the points $A, B, C $. For example, in   
\eqref{A}, \eqref{B}, \eqref{C}, we can use the transformation (as in the $2x2$ case)
$$x_i^*= x_i-x_1, y_i^*=y_i-y_1$$ so we have $$x_1^*= 0,  y_1^*=0$$  which    helps in visualizing the isosceles case, in particular.

\section{Effect of excluding  eigenvalues in the MDS solution}
 When the distance /dissimilarity matrix  is very general, the corresponding matrix   $\bB$ can have some  negative eigenvalues, which  can distort the Euclidean fitted configuration.  We now give   some insight into this possible effect. 
Let $\bD=(\delta_{ij})$ be an $n \times n$ dissimilarity matrix. We are using slightly different notation than in the first section to emphasize that we are working  now on a dissimilarity matrix  and with the fitted  distances  $(d_{ij})$  for the MDS solution .
Suppose as in \eqref{Bspectral} the corresponding matrix $\bB$ has spectral decomposition $\bB = \mathbf{\Gamma} \mathbf{\Lambda} \mathbf{\Gamma}^T$, with the eigenvalues in decreasing order (there is always
at least one zero, and perhaps some negative eigenvalues) where  as in Section 1, $\mathbf{\Lambda}$ is the diagonal matrix with the eigenvalues $\lambda_\ell,  \ell=1,\ldots,n, $ and $\mathbf{\Gamma}$  is the  matrix of the eigenvectors.    
Write
\beq \label{gij}
g_{ij}^{(\ell)} = \lambda_\ell (\gamma_{i,\ell} - \gamma_{j,\ell})^2.
\eeq
Then from \eqref{Xcoord}, the distance between the points $\vect{x}_i^T=({\lambda_1}^{\frac{1}{2}}{\gamma}_{i1},\ldots,{\lambda_n}^{\frac{1}{2}}{\gamma}_{in} ) $ and $\vect{x}_j^T=({\lambda_1}^{\frac{1}{2}}{\gamma}_{j1},\ldots,{\lambda_n}^{\frac{1}{2}}{\gamma}_{jn})$  is given by 
\beq \label{deltaij}
\delta^2_{ij} = \sum_{\ell=1}^n g_{ij}^{(\ell)}                
\eeq
which is an exact identity.  If the MDS solution uses the first $p $  eigenvalues
(assumed to be nonnegative, $p \leq  n $), then the squared Euclidean distances for
this MDS solution are given by
\beq \label{dij}
d^2_{ij} = \sum_{\ell=1}^p g_{ij}^{(\ell)}.                   
\eeq
The difference between \eqref{deltaij} and \eqref{dij} is
\beq \label{diffij}
\delta^2_{ij} - d^2_{ij} = \sum_{\ell=p+1}^n g_{ij}^{(\ell)}
\eeq
and measures the extent at sites $i,j$ to which the MDS solution fails
to recover the starting dissimilarities.

Let us fix 2 sites $i,j$ and consider two mutually exclusive possibilities (of
course more complicated situations can occur):

(a) Suppose $g_{ij}^{(\ell)}$ is near 0 for all $\ell = p+1, \ldots, n$
except for one value $\ell = \ell_1$, say.  Further suppose that
$\lambda_{\ell_1} > 0$.  Then from \eqref{gij} and \eqref{diffij}, we have 
$$\delta^2_{ij} - d^2_{ij} > 0; $$

(b) Suppose $g_{ij}^{(\ell)}$ is near 0 for all $\ell = p+1, \ldots, n$
except for one value $\ell = \ell_2$, say.  Further suppose that
$\lambda_{\ell_2} < 0$. Then again from \eqref{gij} and \eqref{diffij},
$$\delta^2_{ij} - d^2_{ij} < 0 .$$
Hence, if   $g_{ij}$ given by \eqref{gij} is positive (negative), the Euclidean distance will be smaller than (greater than) the dissimilarity. 

We now give a numerical example. 

{\bf Example.}  We   look at the journey times between a selection of 5 rail
stations in Yorkshire (UK)  to understand how the eigenvectors of $\bB$ can help to understand the behaviour of a solution of $\bB$ with some negative eigenvalues.   

There are two rail lines between Leeds and
    York; a fast line with direct trains, and a slow line that stops
    at various intermediate stations including Headingley, Horsforth
    and Harrogate.

    Here the ``journey time'' is defined as the time taken to reach the
    destination station for a passenger who begins a journey at the
    starting station at 12:00 noon.  For example, consider a passenger
    beginning a journey at Leeds station at 12:00.  If the next train for York
    leaves at 12:08 and arrives in York at 12:31, then the journey
    time is 31 minutes (8 minutes waiting in Leeds plus 23 minutes on
    the train).  The times here are taken from a standard weekday
    timetable.

    Table \ref{table:rail} below gives the dissimilarities between all pairs of
    stations, where the dissimilarity between two stations $S1$ and $S2$ is
    defined as the smaller of two times: the journey time from $S1$ to $S2$
    and the journey time from $S2$ to $S1$.  Further the dissimilarity
    between a station and itself is taken to be 0.

    \begin{table}[h]
      \begin{center}
          \caption{ Dissimilarity matrix $\bD$ for train journey
          times between 5 rail stations in Yorkshire.}
     \begin{tabular}{lrrrrr}
        & A & B & C & D & E\\
          (1) A: Leeds   & 0 & 23 & 23 & 53 & 31 \\
          (2) B: Headingley & 23 & 0 & 11 & 34 & 71\\
          (3) C: Horsforth & 23 & 11 & 0 & 34 & 67\\
          (4) D: Harrogate & 53 & 34 & 34 & 0 & 44\\
          (5) E: York & 31 & 71 & 67 & 44 & 0
        \end{tabular}
       \label{table:rail}
      \end{center} \end{table}
  
  \begin{figure}[h]
    \begin{center}
      \includegraphics[width=4in,height=4in]{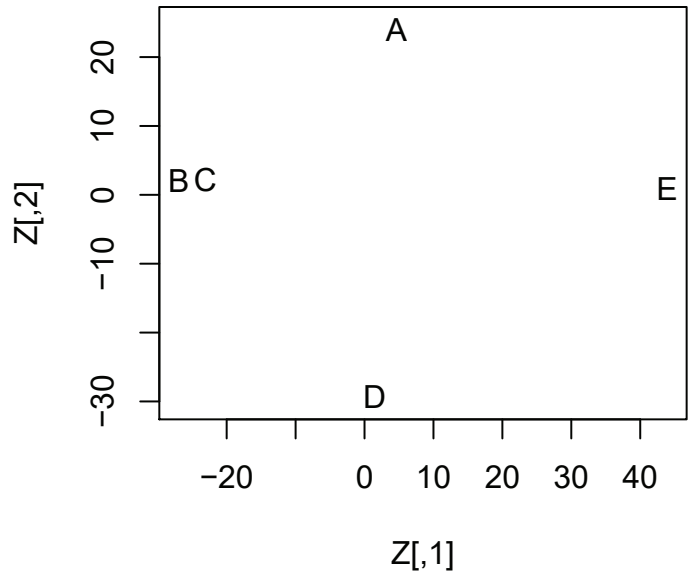}
      \caption{ Two-dimensional MDS solution for train
        journey times between 5 rail stations in Yorkshire. A: Leeds,   B: Headingley,  C: Horsforth, D: Harrogate, E: York.}
      \label{fig:rail}
      \end{center} \end{figure}
\newpage      
The eigenvalues of $\bB$ are  
$$\lambda_1= 3210, \lambda_2= 1439, \lambda_3=61, 
     \lambda_4=0,  
     \lambda_5= -964, $$
 and the corresponding  eigenvectors in $\mathbf{\Gamma}$ are    

\begin{verbatim}
Eigenvectors
      [,1]  [,2]  [,3]  [,4]  [,5]
[1,]  0.08  0.63 -0.06 -0.45  0.63
[2,] -0.48  0.06 -0.66 -0.45 -0.38
[3,] -0.41  0.06  0.75 -0.45 -0.26
[4,]  0.03 -0.77 -0.04 -0.45  0.45
[5,]  0.77  0.02  0.00 -0.45 -0.45

\end{verbatim}
We take our MDS solution to be the two dimensional principal coordinates. Figure \ref{fig:rail} plots   these two dimensional principal coordinates. Obviously
 as seen  by the eigenvalues, $\bD$ is not a distance matrix. Also we can check that  
$$71 = \delta_{25} > \delta_{12}+ \delta_{15}  = 23 + 31 = 54,$$ which violates
      the triangle inequality.

      The eigenvalues are 3210, 1439, 61, 0, -964. The first two are
      considerably larger than the rest in absolute value, suggesting
      the 2D MDS solutions should be
      a good representation.  In particular,
     $\lambda_3=61$ seems negligible,
     $\lambda_4=0$  is an eigenvalue that always appears with eigenvector $\vect{1}_n$,
     $\lambda_5= -964$ is smaller than the first two eigenvalues, but not entirely
      negligible and may cause some distortion in the reconstruction as we now examine.

      Figure 1 shows that the stations lie roughly on a circle (not surprising since
      there are two lines between Leeds and York).  Also, Headingley and
      Horsforth are close together, and Leeds is further from Harrogate than
      from York  in terms of the dissimilarity though geographically Harrogate is nearer to Leeds than York. 
      
      In the MDS solution, the Euclidean distance between Headingley and
    Horsforth is 7.1, which is smaller than the dissimilarity value
    11.  On the other hand, in the MDS solution the Euclidean distance between
    Leeds and York is 45.5, which is larger than the dissimilarity
    value 31.  We can now explain  this behaviour using the spectral decomposition of $\bB$, and using the result derived in this section. Let us now denote the stations A, $\ldots$, E   by $1,\ldots ,5 $ respectively.
 Eigenvector entries for selected stations (and the corresponding eigenvalues 61 and -964 respectively)\\
      \begin{tabular}{lrr}
        Station & $j=3$ & $j=5$ \\
        Headingley (2) & -0.66 & -0.38\\
        Horsforth (3) & 0.75 &  -0.26\\
        absolute difference & {\bf 1.41} & 0.12 \\
                                          \\
        Leeds (1)   &   -0.06 &  0.63\\
        York (5)    &   0.00 & -0.45\\
        absolute difference & 0.06 & {\bf 1.08}
      \end{tabular}
      
Hence, the difference between Headingley and Horsforth is dominated by
      the eigenvector $j=3$ (with positive eigenvalue, 61), whereas the
      difference between Leeds and York is dominated by the
      eigenvector $j=5$ (with negative eigenvalue, -964). In fact, the numerical values of the terms \eqref{gij}  in the difference between the two distances given by  \eqref{diffij}  are  
   $$  g_{23}^{(3)} =  121.3, \quad g_{23}^{(4)}= 0,\quad  g_{23}^{(5)} = -13.9, $$
   and 
    $$  g_{15}^{(3)} = 0.2,\quad  g_{15}^{(4)} = 0,\quad  g_{15}^{(5)} = -1124.4, $$  
    so the dominated contributions $g_{23}^{(3)}$ and $g_{15}^{(5)}$ are clearly seen. This discussion explains  why  in the MDS solution, the Euclidean distance between Headingley (2) and
    Horsforth (3) is smaller than the dissimilarity value
    whereas  the Euclidean distance between
    Leeds (1)  and York (5) is larger than the dissimilarity
    value.
\section{Acknowledgment} We wish to express our thanks to  Wally Gilks and John Kent  for their helpful comments and to the University of Leeds for the Example in Section 4 from an Examination paper. The first author would also like to thank the Leverhulme Trust for the Emeritus Fellowship.
\section{References}
Mardia, K. V., Kent, J. T., and Bibby, J. M. (1979). {\it Multivariate Analysis}. Academic press.

\end{document}